# Protein structure classes — High-dimensional geometry of amino acid composition space revisited

"Too Many Molecules, Too Little Time"

— Ivan Amato, *Chemical and Engineering News*, March 9, 2006

Boryeu Mao

11328 179th Court NE
Redmond, WA 98052
USA

boryeu.mao@gmail.com

Running Title: Geometry of protein structure classes in amino acid composition space

Pages: 50 (including figures and tables)

Figures: 6

Tables: 4

Supplementary material: datasets for protein structure classes derived from SCOPe database

## Abstract

In this study, the distributions of protein structure classes (or folding types) of experimentally determined structures from a legacy dataset and a comprehensive database (SCOP) are modeled precisely with geometric constructs such as convex polytopes in high-dimensional amino acid composition space. This is a follow-up of a previous non-statistical, geometry-motivated modeling of protein structure classes with ellipsoidal models, which is superseded presently in three important respects: (1) as a paradigm shift a descriptive 'distribution model' of experimental data is de-coupled from, and serves as the basis for, a possible future predictive 'domain model' generalizable to proteins in the same structure class for which 3-dimensional structures have yet to be determined experimentally, (2) the geometric and analytic characteristics of class distributions are obtained via exact computational geometry calculations, and (3) the full data from a comprehensive database are included in such calculations, eschewing training set selection and biases. In contrast to statistical inference and machine-learning approaches, the analytical, non-statistical geometry models of protein class distributions demonstrated in this study furnish complete and precise information on their size and relative disposition in the high-dimensional space (vis-à-vis any overlaps leading to class ambiguity and classification limits). Intended principally to be an accurate and summary description of the complex relationships between amino acid composition and protein structure classes, and suitably as a basis for predictive modeling where permissible, the results suggest that pen-ultimately they may be useful adjuncts for validating sequence-based protein structure predictions and contribute to theoretical and fundamental understanding of secondary structure formation and protein folding, demonstrating the role of high dimensional amino acid composition space in protein studies.

## Keywords

amino acid composition of proteins; secondary structure of protein conformation; protein structure classes and folding types; geometry of high dimensional amino acid composition space; computational geometry; database of experimentally determined protein structures.

## Introduction

The sequence of amino acid residues of a polypeptide chain, from an 'alphabet' of 20 naturally occurring amino acids, constitutes the primary structure of a protein molecule. Short-range interactions among sidechains of neighboring amino acid residues along the polypeptide backbone lead to the local formation of secondary structure elements such as α-helices and β-strands that further organize into tertiary motifs through long-range interactions among residues far apart along the polypeptide chain but brought to spatial proximity in the overall folded three-dimensional conformation. Each of these tertiary motifs, also called protein structure domains, typically consists of a hydrophobic core with abundant contacts among elements within but weak or little interactions without. Three-dimensional protein structures at atomic resolution solved by x-ray crystallography or NMR spectroscopy show clearly recognized patterns in which α-helices and/or β-strands are well packed into structural domains that have been categorized into different folding types, or classes, according to the content of predominant secondary structures: the class of α or β if there are significant numbers of amino acid residues in either secondary conformation, or mixed classes of α/β and α+β if both are present [1,2]. These protein structure classes play a central role in understanding protein structure and function and evolutionary biology in general on the one hand, and physico-chemical forces driving the formation of secondary structures on the other. While the amino acid sequence (namely the linear order of amino acid residues along the polypeptide chain) is the primary determining factor in the formation of any local secondary structure element, the composition of amino acid residues without the order information is correlated with secondary structure content in protein structure [3,4], and the 20-component composition vectors span a quasi-continuum space ideal for geometric investigation of the relationship between protein classes and amino acid composition [5,6].

From distribution patterns of amino acid compositions of a curated basis set of protein structures then available, and a set of threshold criteria (of more than 15% of the amino acid residues in α helix and less than 10% β strand for type α, less than 15% in α helix and more than 10% β strand for type β, and above both thresholds without dominance of either for α,β mixed types), Nakashima et al. [5] showed that (1) the average of amino acid compositions of the α class is distinctively different from that of β, and (2) protein structures of the different folding types are located in distinctive regions in the amino acid composition space each centered on the average of a given class with a root-mean-square (rms) radius. With the premise that the distribution of composition vectors of a given class is grouped together in the amino acid composition space, it was hypothesized that proteins meeting the threshold criteria for the type but outside the representative set, regardless of the specific content levels of the secondary structure [3,4], can be found in the corresponding region of the amino acid composition space, the region constituting the domain of influence for the forces underlying the formation of the secondary structures found in the folding type. Rules for determining the overall folding type of a protein molecule were then established based on the distance relationships of its amino acid composition vector relative to the class domains in the composition space.

In a subsequent study with similar geometric and non-statistical motivations [6], the product moment matrix of the amino acid compositions of the legacy basis set proteins of Nakashima, Nishikawa and Ooi (henceforth referred to as 'NNO basis set') in a given class is diagonalized for deducing parameters of an ellipsoid in the composition space as a model of the class, and a modified, 'elliptically-scaled', distance measure was derived for classification criterion. In essence the two studies share the common idea of deriving a generally applicable distance measure for classification from the distribution of representative proteins of protein structure classes in the amino acid composition space, but arrive at different distance measures due to different protocols in constructing the amino acid composition space itself. In Nakashima et al. [5], the components of the amino acid composition vector of a protein are calculated for the 20 naturally occurring amino acids relative to the overall averages and then normalized by their respective standard deviations, a procedure in which the relative scale of the 20 coordinate axes of the vector space, and consequently its metric, are effectively changed by the dataset-dependent normalization. In contrast, in Mao et al. [6] the plain amino acid compositions are used directly without normalization, for which the composition space is therefore only 19-dimensional, one degree of freedom fewer due to the single constraint that the 20 components sum to 1. Concomitantly, without effectively re-scaling the coordinate axes of the vector space, an ellipsoid, a high-dimensional equivalent of an ellipse or hyper-ellipse, is derived from the ***p***roduct ***m***oment ***m***atrix of a distribution and its ***d***iagonalization (***pmmd*** ellipsoid), which supplies geometric parameters for the 'elliptically-scaled' distance measure to programmatically account for the shape, size and the orientation of the distribution. The overall objective is to derive class models in the amino acid composition space from the basis set of protein structures, models that would be applicable to other proteins that the basis set may ostensibly be representing. Overlaps among the distribution models of folding types were recognized in both studies, and the intention of either an implicit spherical model [5] or more explicitly the 19-dimensional pmmd ellipsoid [6] was that relevant properties beyond size and shape, such as the overlap between distributions, may also be quantified analytically or computationally.

Although the overall objective of deriving class models is putatively met with pmmd ellipsoidal geometry and the associated distance measure, two dissatisfying implications were clear and recognized at the outset: (a) given the conclusion that the high dimensional amino acid composition space will be inherently a sparsely inhabited space [6], a necessary requirement for improving folding type classification is a substantial increase in protein structural data of the basis set, and (b) the pmmd ellipsoid is a compound model with parameters that are part geometric information of the distribution of basis set proteins and part projection for proteins that basis set proteins ostensibly represent, parts that might be unbundled for more robust and effective modeling of the protein folding classes. Advancing in these directions would await the emergence and progress of protein structure databases [7-9], as well as developments in computational geometry algorithms and methods [10-15]. In re-visiting the high dimensional geometry of amino acid composition and protein folding types, in Section 1 of Results we argue for deferring the pmmd approach, in favor of uncoupling modeling *data distributions* of *specific proteins* in the basis set from the conceptually and procedurally separable modeling of *class domains*, for *all proteins* which proteins in the basis set are intended to be representatives of. In this 'divide and conquer' approach, the distribution modeling would precede, and serve as the basis for, the more general and arguably more complex domain modeling. Continuing with plain amino acid compositions (i.e. without normalization), data distributions of protein folding classes in NNO basis set are then modeled with alternative and more suitable geometric models of minimum-volume enclosing ellipsoid [10] and convex polytope [12,14]. In Section 2, for improving the representation of protein classes in a basis set, protein data from extended SCOP structural database (SCOPe) [8,9] are parsed for data distributions of a comprehensive basis set. Models of structure classes are compared with those from NNO basis set in their geometry. Relevant questions to ask of the data distribution models are: location in the composition space, size and shape of the occupied space, and relative geometric disposition among classes, which may be (a) overlapping (inherently ambiguous and unresolvable structure classes for certain amino acid compositions, the extent of which places a limit on how well protein structures could be classified [6,16,18] and may suggest alternative distribution models), or (b) complete separation (clearcut and straightforward structure classification based on amino acid composition).

Beyond protein structure classes, amino acid composition and related attributes had been found to be relevant to interesting aspects of protein structure, function and evolution [19-22]. A fundamental understanding of the relationship between protein folding types and amino acid composition exemplified in the geometry of the high-dimensional space nevertheless would help ultimately delineate the physico-chemical forces underlying and influencing secondary structure formation [4] and overall folding into the three-dimensional protein conformation [23].

For facilitating the presentation of materials, set notations are used throughout the text, and relevant routines and procedures are documented in pseudo code and collected in the Methods section. Datasets for the study as well as actual codes are available upon request. Data processing and computations were carried out on a Linux virtual machine (Debian 12 operating system) hosted in Qubes OS hypervisor (release 4.2.3) running on a Dell 3505 computer, with dual-core Ryzen 5 processor and 16 megabytes of memory.

# Results

## 1. Geometric Representation

In the pmmd approach for modeling the relationship between protein folding class and amino acid composition [6, and also 17], it was hypothesized that the envelope of pmmd ellipsoid prescribes the domain boundary of the class in the amino acid composition space[1] within which the elliptically-scaled distance is a useful measure for classifying proteins to that class, a boundary where physico-chemical forces influencing the secondary structure formation in the folding type would be diminished and the amino acid composition no longer as relevant for the folding type. In order for covering proteins outside of the basis set of the class, the hypothesized domain boundary might be expected to expand from the immediate neighborhood of data distribution (as shown for the pmmd ellipsoid for the 2D illustration in Figure 1). Hypothesis testing and refinement of the pmmd ellipsoid or similar models for class domains would be complex due to the lack of a full mechanistic understanding of the physical and chemical forces underlying the formation of secondary structures. A more effective alternative would be to re-cast the structure classification problem, and uncouple modeling the data distribution of specific proteins of a class in the basis set from domain modeling appropriate for all proteins of that class, with separate hypothesis testing and validation as necessary; that is, deferring compound model such as pmmd ellipsoid in favor of a *descriptive model* of data distribution of basis set proteins and a conceptually and procedurally separate *domain model*.

Figure 1

**1** *Amino acid compositions are fractions calculated from integer counts of individual amino acids and their sum. Accordingly, the amino acid composition space is the space of rational numbers, denoted as $\mathbb{Q}^n$, spanned by 20 axes, each of discrete rational numbers in the interval [0,1] and for a naturally occurring amino acid, subject to the constraint that the 20 components sum to 1. Protein structure classes occupy parts of $\mathbb{Q}^n$ that are compositions realizable by proteins of finite lengths with length distributions genomically specified in proteomes.*

Following the class designation of 1 to 5 [6], data distributions of classes (or folding type FT) for NNO basis set are denoted as follows:

$\underline{\text{class}}\langle i \rangle := \{$ amino acid composition vector $\bar{v}$ of protein | protein $\in$ FT(i) $\}$, where

$i \in \{1,2,3,4,5\}$, FT(i) $\in \{\alpha, \beta, \alpha/\beta, \alpha+\beta, \text{irregular}\}$,

with the cardinality of the classes, denoted with collective term **I** := {1,2,3,4,5}:

$N(\underline{\text{class}}\langle \mathbf{I} \rangle) := \{29,34,39,26,4\}$.

Formal definitions of sets and functions and procedures for geometric calculations, as well as symbols, notations are found in Methods section.

For the purpose of a suitable and more precise descriptive model of class distribution, the 'loosely fitting' pmmd ellipsoid conceivably could be scaled inward radially such that its envelope comes in contact with the outer-most data points in the distribution (i.e. vertices in Figure 1). This however will not be the smallest such ellipsoid except in special cases (when the distribution is symmetrical, for example). Instead, for a strictly smallest and hence more precise descriptive model covering a distribution, a well-known algorithm is the ***m***inimum ***v***olume ***e***nclosing ellipsoid (***mve*** ellipsoid) [10], and the python code for arbitrary dimensions is available and published in github.com [11]. Results of pmmd ellipsoid *P*(**J**), reproduced for backward reference and for comparison, and mve ellipsoid *M*(**J**) for **J** := {1,2,3,4} are shown in Table I. Rather than floating point numbers in percentage unit before [6], components of the amino acid composition vector are now fractions in the rational number form, and thus the volume in 19D [6,24 and Methods] smaller by a multiplicative factor of $10^{-38}$, with a slightly higher numerical precision due to computing in rational number arithmetic.

Table I

As illustrated in Figure 1, an mve ellipsoid is a more precise representation of a point distribution than pmmd ellipsoid due to its reduced envelope, but an even more and ultimately the most concise geometric form is that of convex polytope, essentially a shrink-wrap around the point set, and an extensively researched and active field of computational geometry [12,14]. Unlike pmmd ellipsoid, both mve ellipsoid and polytope are constitutionally defined by the boundary of the distribution; any internal structure (e.g. the 7 interior points in Figure 1) or any duplicate data points are 'redundant' in that they do not change the geometric object. A convex polytope, as exemplified by the polygon, a 2D polytope, shown in Figure 1, is the region in real space $\mathbb{R}^{n}$[2] defined by its set of vertices, or equivalently bounded by a set of facets (i.e. line segments in 2D between neighboring pairs of vertices) that collectively form the envelope covering the point set. Lines extending from the facets (in long dashes) are half planes which are duals of vertices defining the polytope [12,14]. For the purpose of this study, three groups of computational geometry calculations are defined for a polytope model *T* of a distribution D (Methods II(a)): (1) a redundancy test, redund(D), for identifying points that are 'redundant' in its definition, and vertex(D), for the remainder in D that are non-redundant and essential; (2) inter-conversions between dual representations $T_{\mathrm{v}}$ and $T_{\mathrm{h}}$ of polytope in vertices and in half-planes respectively; and (3) from $T_{\mathrm{v}}$, a computation of the content of a polytope ('area' for 2D polygons, 'volume' for 3D polyhedra, and generalized to 'hyper-volume' for higher dimensional polytopes). As the computing resources required for polytope calculations scale combinatorically with the number of

---

**2** *It is worth noting that the geometry of rational coordinate space $\mathbb{Q}^{n}$ is limited; for example, the distance is in general real and not rational, and certain geometric shapes cannot be accommodated in this space. For facilitating general geometric discussions (e.g. Table I), the amino acid composition space is implicitly treated as embedded in the real coordinate space and protein structure classes are modeled as geometric objects in $\mathbb{R}^{n}$.*

points in the distribution, mve ellipsoid is a useful surrogate where polytope calculations become excessively expensive. The basic concepts of computational geometry calculations relevant to this work are outlined and illustrated schematically in Figures 1 and 2. Readers are referred to computational geometry texts and resources for further details as needed.

Geometric parameters computed for the four main folding types class⟨**J**⟩ are shown in Table I. Compared to the 2D example in Figure 1, the volume reduction from pmmd ellipsoid to mve ellipsoid and to polytope is more pronounced in 19D. For NNO basis set, all redund(class⟨**J**⟩) is null, or equivalently, vertex(class⟨**J**⟩) and class⟨**J**⟩ are identical; that is, all representative proteins are vertices of their respective polytopes, without any redundancy or interior points.

Figure 2

For two distributions $\underline{D}_1$ and $\underline{D}_2$, any intersection of their polytope representations[3] that might exist can be readily evaluated in a relatively straightforward vertex enumeration $\underline{O}_p(\underline{D}_1, \underline{D}_2)$ (Methods II.4), since the region of intersection is itself a polytope, as shown schematically in orange in Figure 2. For polytope in high dimensional space of even only a few vertices, the combinatorial increase in the number of half-planes (Table I) would require significant dedicated computer resources (processor speed, core size and disk storage) for direct computation of $\underline{O}_p$. As an economizing and practical alternative, provisory overlap information can be deduced from examining the following types of discrete overlap (Methods II(b)), in the order of decreasing proximity of $\underline{D}_1$ and $\underline{D}_2$: full and partial centroid overlap $\underline{O}_c(\underline{D}_1|\underline{D}_2)$, distribution overlap $\underline{O}_d(\underline{D}_1|\underline{D}_2)$, as illustrated in Figure 2.

***3*** *The polytope intersection between two distributions, as defined in Methods II.4-6, is distinct from the intersection of the distributions themselves, which refers to the set of common members, that is, amino acid compositions that appear in both distributions. This common set can be identified without polytope computations, and may be considered as a more fundamental overlap than polytope overlaps. No common amino acid compositions exist among classes in NNO dataset.*

Among $\underline{\text{class}}\langle\mathbf{J}\rangle$, all $\underline{O}_{\text{c}}$ and $\underline{O}_{\text{d}}$ are null. However the distributions may not be completely separated from one another since there may still be facet overlaps (e.g. two prisms in contact at edges and faces) that must be ultimately evaluated with polytope $\underline{O}_{\text{p}}$. For classes 1 and 4 for example, with the smallest combined set of half-planes (Table I, row 7), the preliminary results of $\underline{O}_{\text{p}}(\underline{\text{class}}\langle 1\rangle,\underline{\text{class}}\langle 4\rangle)$, a long vertex enumeration calculation courtesy of David Avis, show that the two distributions do indeed overlap ($\underline{O}_{\text{p}}(\underline{\text{class}}\langle 1\rangle,\underline{\text{class}}\langle 4\rangle)$ is not null). Other bipartite combinations are larger in size and their feasibility better resolved by linear programming (Skip Jordan, Otaru University of Commerce, Japan). The results confirm that $\underline{\text{class}}\langle 1\rangle$ and $\underline{\text{class}}\langle 4\rangle$ overlap and show that $\underline{\text{class}}\langle 1\rangle$ and $\underline{\text{class}}\langle 2\rangle$ do not. To summarize, class distributions of NNO basis set proteins are intersecting only on the periphery, implying that overlaps of pmmd ellipsoids observed previously [6] are overlaps of hypothesized class domains.

Given that redund($\underline{\text{class}}\langle\mathbf{J}\rangle$) is null for all **J**, proteins in NNO basis set define only the polytope boundary for the classes but are too few for any internal occupancy in the high dimensional space, arguably an indication that they are poor overall representations of the folding classes: the distributions are either 'small' and 'thin', or otherwise 'superficial' without internal quasi-continuum, to adequately support claiming and hypothesizing that class domains do represent the folding types more widely. These considerations point to a comprehensive dataset for remedying the poor data representation (i.e. a 'whole-database' approach), in preference to training data that are revised, augmented, or otherwise expanded from NNO basis set for improving representativity of the classes, the traditional machine-learning approach (for example [16,18,25]) in which risks of potential inherent or hidden bias of training set selections are mitigated with statistical inferencing.

## 2. Protein Database

CATH [7] and SCOP [8,9], two major comprehensive, hierarchical protein structure databases were initially developed and available in mid- to late-1990's and continuously evolving since. Both databases collect 3D protein structures from Protein Data Bank (PDB), https://www.rcsb.org, a repository of experimental structure data from x-ray crystallography or NMR spectroscopy, and in conjunction with genetic and sequence information derive a classification hierarchy of evolutionary and functional relationships among proteins. In both databases, the classification at the top level identifies folding types according to the predominant secondary structure content of α helices and/or β strands. Structures with mixed α and β are grouped into a single class in CATH [26], but separately into two classes α/β and α+β in SCOP [27] as in classification studies before [5,6] and thus more suited as an expanded basis set for protein structure classes[4]. In the 7-level SCOP hierarchy (Figure 3), the major classification is encoded in an alpha-numeric ***S***COP ***c***oncise ***c***lassification ***s***tring (***sccs***), in the form of **cl**.**cf**.**sf**.**fa** for the top four levels of **Class**, **Fold**, **Superfamily**, and **Family** respectively. At the 7th and the lowest level of SCOP hierarchy are protein structure domains **px**, regions in the 3D protein conformation within which amino acid residues are well-packed into a structural core via optimal and stabilizing inter-residue interactions. The **px** entries, being the unit of classification in SCOP [27] and identified by a 7-character **sid** string containing the associated PDB ID, correspond to members of NNO basis set. Evolutionarily related proteins and their structure domains are grouped into families at level **fa** [28], which then are organized into higher levels up to **Class** (level **cl**). Many structure domains are single-domain proteins of single polypeptide chains, while others are multiple domains

***4*** *For the purpose of expanding the basis set for protein classes, experimental databases remain the 'gold standard', and prediction databases such as AlphaFold (of millions of structures) [23] may be useful complement if or when fully automated class assignment protocols are available (see Discussion and Summary).*

within certain long polypeptide chains, or multi-chain domains consisting of more than one polypeptide chain.

Figure 3

SCOPe parseable files dir.cla.scope.txt and dir.com.scope.txt [29], and its ASTRAL sequence resource file astral-scopedom-seqres-gd-all-2.08-stable.fa [30] are downloaded for version 2.08 of the database with 2023-01-06 updates. A small number of amino acid sequences not in the ASTRAL file are fetched from SCOP individually [31]. In addition to the four major SCOP classes, a, b, c, and d, corresponding to α, β, α/β and α+β respectively, also included in this study is class g, a group of small proteins without significant secondary structures but stabilized by disulfide bonds or metal ion chelation. Pre-processing of SCOPe data and statistics are detailed in Table II-a and II-b respectively. Data distributions for the folding classes (Table II-b, row 4) are denoted as follows:

Table II

$$\underline{\text{class}}\langle k_0\rangle := \{\text{ amino acid composition vector of } \mathbf{px} \mid \text{sid of } \mathbf{px} \in \text{SCOP class } k\ \}, \text{ where}$$

$$k \in \{a,b,c,d,g\},\ \text{SCOP folding-type}(k) \in \{\alpha, \beta, \alpha/\beta, \alpha{+}\beta, \text{small proteins}\}$$

$$N(\underline{\text{class}}\langle k_0\rangle) \in \{32957, 55705, 53404, 56681, 6007\}$$ [5]

with collective terms $\underline{\text{class}}\langle \mathbf{K}_0\rangle$, $\mathbf{K}_0$ and $\mathbf{L}_0$ defined similarly to those for NNO basis set. With $N(\underline{\text{class}}\langle \mathbf{K}_0\rangle)$ being 3 orders of magnitude larger than NNO basis set, full computational geometry calculations would require significant hardware resources in processor speed, core memory, and disk storage[6]. For reducing resource needs, however, we make use of the hierarchical data structure and bootstrap[7] from a higher SCOP level; for example, for distributions of folding classes $k \in \{a,b,c,d,g\}$ at the **Protein** level (**dm**),

---

5 *When replicates within a class are removed, these counts are reduced to: 10386, 15139, 14880, 14441, 2503, essentially within the same respective order of magnitude.*

6 *To paraphrase Ivan Amato: "Too Many **Protein** Molecules, Too Little **Computer** Time."*

7 *Bootstrapping, as in 'Leveraging a small, initial effort', here unrelated to bootstrapping operations of resampling in statistics.*

protein⟨u⟩ := { amino acid composition vector of **px** | sid of **px** ∈ SCOP sunid-dm u },

where sid is the 7-character alphanumeric identifier of protein domain **px**, and sunid-dm is the SCOP internal unique numeric identifier **sunid** (https://scop.berkeley.edu/help/ver=2.08) for members of the hierarchy at the sixth level **dm** (Figure 3). With N({protein⟨u⟩ | u ∈ class k}) of 2285,2333,2773,2966,743 respectively (Table II-b, row 5), which are lower than class⟨$\mathbf{K}_0$⟩ (Table II-b, row 4), though still 2 orders of magnitude larger than class⟨**I**⟩ (Table-I, row 6), with combinatorically yet higher counts expected for half-planes. The situation becomes more manageable at the **Family** level (**fa**):

family⟨s⟩ := { amino acid composition vector of **px** | sid of **px** ∈ SCOP sccs-fa s },

with N({family⟨s⟩ | s ∈ k.∗.∗.∗ }) ∈ {893,766,807,1093,230}, for k ∈ **K** (Table II-b, row 6). Class distributions are derived from family⟨s⟩ as follows (Methods I, II):

Family⟨s⟩ := vertex(family⟨s⟩) (when vertex(family⟨k⟩) ≠ family⟨k⟩)

class⟨k⟩ := {$\overline{\text{centroid}}$(Family⟨s⟩) | SCOP sccs-fa s ∈ k.∗.∗.∗ }

Class⟨k⟩ := vertex(class⟨k⟩) (when vertex(class⟨k⟩) ≠ class⟨k⟩)

$\overline{\text{centroid}}$⟨k⟩ := mean(Class⟨k⟩).

For folding type α, for example, distribution class⟨$a_0$⟩, of 32957 **px**'s, is refactored hierarchically into class⟨a⟩ of the centroids of 893 **fa**'s, with N(family⟨s⟩) ranging from 1 to 2750 and N(Family⟨s⟩) from 1 to 528 (Table II-b, rows 7 and 8 respectively). Figure 2 schematically shows the relationship between $T_v$(Class⟨k⟩) (polytope in blue), two of its Family⟨s⟩ (marked by blue x), and partial $T_v$(Class⟨$k_0$⟩). (short dashed lines in blue). The choice of **Family** level as an effective unit of classification for protein folding types is also justified by its unique position in the hierarchy of protein structure, function and evolution [2,27]. The wide range of values for N(family⟨s⟩) (Table II-b) reflects the great diversity in protein families. For relatively small Family⟨s⟩ (N(Family⟨s⟩)<51), $V_T$ are readily calculated and shown to correlate with $V_M$ over 30 orders of magnitude (Figure 4), in essence a demonstration that an ellipsoid can mve-cover many different polytopes. The empirical correlation relationships are useful for estimating $V_T$ of polytopes with high vertex count, when $V_M$ is more readily calculated and the value falls within range of the observed correlation regime appropriate for extrapolation.

Figure 4

The dataset size of class⟨**K**⟩ (Table III-a) is significantly larger than NNO basis set, with approximately 20- to 30-fold increase from N(class⟨**I**⟩); for folding type α for example, N(class⟨a⟩) is 893, with 824 vertices (N(Class⟨a⟩)). Also, significantly larger than class⟨5⟩, class⟨g⟩ is now full rank. The comparison of distinctively different amino acid compositions of folding types α and β for NNO basis set (Figure 5, top panel for class⟨1⟩ and class⟨2⟩, and comparable to Figure 1 in [5]), shows that, geometrically speaking, $\overline{\text{centroid}}$⟨1⟩ and $\overline{\text{centroid}}$⟨2⟩ are located respectively in quadrants II and IV of the coordinate plane the origin of which is the overall mean of class⟨**I**⟩ and the two axes parallel to the collective axes of amino acids {K,A,E,M,H,L,D} and {N,Y,V,P,G,C,T,S} respectively, the two groups of amino acids favoring α helices (amino acid letter codes in blue in Figure 5) and β strands (letter codes in orange).

Figure 5

Similar to class⟨1⟩ and class⟨2⟩, the amino acid compositions of Class⟨a⟩ and Class⟨b⟩ are shown to be distinctively different (Figure 5, bottom panel). The re-ordering of amino acids within the left and the right halves (primarily L,E,A,K,M for α and N,S,P,T,V,G for β), however shows that directional vectors of types α and β are re-oriented from those for class⟨1⟩ and class⟨2⟩, but remain essentially in the respective quadrants, suggesting that the expansion of the legacy basis set proteins to a whole database refines but does not substantively alter the geometric relationship between folding types α and β. Similar collective amino acid compositions were previously noted to correlate with α helix content estimated from optical rotary dispersion measurements predating modern day x-ray diffraction and NMR spectroscopy [3,4].

Table III

Compared to NNO basis set, data distributions of class⟨**L**⟩ occupy wider regions in the amino acid composition space, i.e. $V_M(\mathbf{L}) > V_M(\mathbf{J})$ (Table III-a, row 2 vs Table I, row 3), and $V_{T\text{-est}}(\mathbf{L}) > V_T(\mathbf{J})$ (last rows, Table III-a vs Table I). $M\langle\mathbf{L}\rangle$ are less eccentric than $M\langle\mathbf{J}\rangle$. In contrast to NNO dataset, redund(**K**) is not null (Table III-b, row 1), e.g. of the 893 families of class⟨a⟩, 69 are located in the interior of $T\langle a\rangle$. Except for class⟨g⟩, the interior subsets themselves are single-layer polytopes, and nearly concentric with their respective $T\langle\mathbf{K}\rangle$. These results show that, by virtue of both outward and inward expansion of the distributions, SCOPe data improves the overall representation of the protein folding classes. As illustrated in Figure 2, the envelopes of $T_v$(Class⟨$\mathbf{K}_0$⟩) (short dashed lines in blue) are expanded from the penultimate $T_v$(Class⟨**K**⟩) (solid lines in blue). The average size of **Family** distributions (marked by x in Figure 2),

$$\text{mean}(\{\dim_M(\underline{\text{Family}}\langle s\rangle) \mid s \in \underline{\text{Class}}\langle k\rangle\})$$

are {0.0183, 0.0183, 0.0135, 0.0173} respectively for Class⟨**L**⟩, equivalent to increases of {12.6%, 13.2%, 14.1%, 12.8%} (relative to $\dim_M$ (Class⟨**L**⟩), Table III-a, row 4) of V(Class⟨**L**⟩) for the estimated V(Class⟨$\mathbf{L}_0$⟩).

Table IV

As a result of the larger volume of class⟨**L**⟩ compared to class⟨**J**⟩, and perhaps secondarily re-orientation of the distributions (Figure 5), there are now distribution overlaps[8] among the classes (Table IV-a), and centroid overlaps with Class⟨c⟩ or Class⟨d⟩ but not between Class⟨a⟩ and Class⟨b⟩ (Table IV-b). Note however that there are no overlaps of either type between any of class⟨**L**⟩ with class⟨g⟩. For each class⟨k⟩, the distribution overlaps with all other classes (last row, Table IV-a) are not redundant to the respective inner polytopes; they constitute a fraction of its total, namely, $N(\{\underline{O}_d(\underline{\text{class}}\langle k\rangle \mid \underline{\text{Class}}\langle k'\rangle) \mid k' \neq k\}) < N(\underline{\text{class}}\langle k\rangle)$, or, in other words, the remainder of class⟨k⟩ are located in non-overlapping part(s) of $T\langle k\rangle$ unique to the class. In particular, $T\langle c\rangle$ and $T\langle d\rangle$ (and estimated $T\langle c_0\rangle$ and $T\langle d_0\rangle$), though clustered close together, do not completely overlap with each other, in support of two separate classes for α/β and α+β as in SCOP [27,32], rather than a combined mixed α and β class as in CATH [7,26]. The eventual complete calculation for full distributions $\underline{O}_p(\underline{\text{class}}\langle c_0\rangle, \underline{\text{class}}\langle d_0\rangle)$ will determine if class α/β falls completely within, and therefore a subset of α+β.

Centroid-to-centroid distances among class⟨**L**⟩ are shown in Table IV-b (rows 1-5): $T\langle a\rangle$ and $T\langle b\rangle$ are farthest apart, at a distance of 0.0696, $T\langle c\rangle$ and $T\langle d\rangle$ closest together at a distance of 0.0171, while $T\langle g\rangle$ is located far away, at distances greater than 0.1 from, and not overlapping with, $T\langle \mathbf{L}\rangle$. Of the four major folding types, $T\langle d\rangle$ is located virtually at the center $\overline{\text{centroid}}(\overline{\text{centroid}}\langle \mathbf{L}\rangle)$ (Table IV-b, row 6). At equal distances from $T\langle d\rangle$ that sum up to approximately the $T\langle a\rangle$-$T\langle b\rangle$ distance, $T\langle a\rangle$ and $T\langle b\rangle$ are located diametrically from the group center, consistent with Figure 5. Given full bipartite

**8** *As noted before (footnote **3**), the distribution overlaps here refer to overlaps between polytope representations of distributions defined in Methods II.4-6. For two distributions as sets, the intersection refers to common set members. Within a distribution, common set members are replicates. In SCOP hierarchy, there are many replicate amino acid compositions within groups at **Family**, **Protein**, and **Species** levels, which are treated as 'redundant' in polytope computations. At higher levels, among classes a,b,c,d,g, there are replicates only at **Superfamily** level, for a total of 3 sets: 2 in superfamily "immunoglobin" (b.1.1), between families b.1.1.1 and b.1.1.2, and 1 in superfamily "nucleic acid-binding protein" (b.40.4), between b.40.4.5 and b.40.4.11. More importantly, there are no common amino acid compositions at the **Class** level, i.e. no intersection ('intrinsic overlap') among* $\underline{\text{class}}\langle \boldsymbol{K_0}\rangle$.

centroid overlaps among $T\langle\mathrm{b}\rangle$, $T\langle\mathrm{c}\rangle$ and $T\langle\mathrm{d}\rangle$, and nearly full bipartite centroid overlaps among $T\langle\mathrm{a}\rangle$, $T\langle\mathrm{c}\rangle$ and $T\langle\mathrm{d}\rangle$, (the only exception being $T\langle\mathrm{a}\rangle$ and $T\langle\mathrm{c}\rangle$), classes b,c,d are clustered slightly more tightly than classes a,c,d, with a triangular area of the centroids of the b,c,d-cluster, or $\mathrm{V}_T(\{\overline{\mathrm{centroid}}\langle\mathrm{k}\rangle \mid \mathrm{k} \in \{\mathrm{b,c,d}\}\})$, of 6.130e-6, compared to 7.316e-6 for $\mathrm{V}_T$ of the a,c,d-cluster. Together these geometric parameters are summarized in the following clustering relationship of $T\langle\mathbf{K}\rangle$ in order of decreasing proximity (and schematically illustrated in Figure 6):

$$T\langle\mathrm{c}\rangle,T\langle\mathrm{d}\rangle > T\langle\mathrm{b}\rangle,T\langle\mathrm{c}\rangle,T\langle\mathrm{d}\rangle \gtrsim T\langle\mathrm{a}\rangle,T\langle\mathrm{c}\rangle,T\langle\mathrm{d}\rangle > T\langle\mathrm{a}\rangle,T\langle\mathrm{b}\rangle > T\langle\mathrm{g}\rangle,T\langle\mathbf{L}\rangle$$

For class g, a group of proteins with little or no α or β secondary structures, $T\langle\mathrm{g}\rangle$ is located in the amino acid composition space far away from $T\langle\mathbf{L}\rangle$ (Table IV-a) and also at a large distance to $\overline{\mathrm{centroid}}(\overline{\mathrm{centroid}}\langle\mathbf{L}\rangle)$ (Table IV-b), that is, amino acid compositions of proteins in class g are significantly different from those with α or β secondary structures and thus as a group is a proper negative control for associating amino acid composition in the region around $\underline{\mathrm{class}}\langle\mathbf{L}\rangle$ with α helix or β strand secondary structure formation.

Figure 6

Relative to SCOPe release 2.07, database additions and updates in release 2.08 (Table II) did not substantively alter the geometric dispositions of the folding classes (Table IV-b, row 8).

Compared to distances among $T\langle\mathbf{L}\rangle$ for SCOPe dataset (Table IV-b, rows 1-6), the corresponding values for $T\langle\mathbf{J}\rangle$ in NNO basis set are larger, which is consistent with a slightly exaggerated difference between folding types α and β (Figure 5), and quantified as: $\mathrm{V}_T(\overline{\mathrm{centroid}}\langle\mathbf{J}\rangle) > \mathrm{V}_T(\overline{\mathrm{centroid}}\langle\mathbf{L}\rangle)$, at 1.379e-7 and 2.783e-9 respectively. Individually and in reference to class distributions in SCOPe on the other hand, NNO basis set proteins populate the domains only sparsely and non-uniformly (small N(class⟨**J**⟩)), covering only small sub-regions (smaller $V_T$ or $V_M$ than class⟨**L**⟩), and off-centered (Table IV-b, row 7).

## Discussion and Summary

For classification of protein folding types, the distributions of specific proteins in the legacy NNO basis set of Nakashima et al. [5] in the amino acid composition space are re-modeled with computational geometry constructs. Although there is little or no intersection among the distributions (i.e., distinctively different amino acid compositions for the four main structure classes), the basis set distributions are too small, in both set cardinality and volume content, and inherently limited to be fully representative of the folding classes. SCOPe datasets on the other hand not only substantively expand the coverage of the folding type domains for the four main folding types but also natively furnish a fifth group, of small proteins with little or no α helix or β strand secondary structures, as a proper negative control. For data distributions of SCOPe basis set, folding type α+β is located near the center of the four main folding types in the amino acid composition space, clustered closely with the most compact distribution of class α/β and flanked by classes α and β. Computational geometry modeling of data distributions from the SCOP database overcomes limitations of the smaller NNO basis set, and also addresses long-standing open questions in the previous study [6]. The complex amino acid composition-structure class relationship shown in Table IV suggests that the inherent limitation placed on classification accuracy may be more nuanced but perhaps somewhat less pessimistic than that derived from machine learning approaches (e.g. 60% correct prediction for upper limit [16]).

The geometry of the distributions of protein classes in the high dimensional space accurately demonstrated the complex relationship between folding types and amino acid compositions, knowledge that would contribute to developing predictive models and could be a useful adjunct for validating sequence-based protein structure prediction (AlphaFold and others). The geometry-motivated whole-database approach may also contribute to the fundamental and mechanistic understanding of the formation of secondary structures in protein conformation (the four main classes vs. the class of 'small proteins'), and possibly even more broadly to proteomics (folded vs. disordered protein molecules [33-35], for example).

SCOPe undergoes periodic updates, as newly released protein structures in PDB databank are incorporated into the database. There are only small increments in the dataset from SCOPe release 2.07 to 2.08 (Table II), with minimum effects on the overall geometry of the folding classes (Table IV-b, row 8), a trend to be expected for a maturing database. Any addition of unusual structures in a particular release may be detected by continually monitoring the geometry of class distributions.

In addition to periodic updating from SCOP releases, further work of interest on data distributions of protein classes includes the following:

(1) with sufficient computing resources, full computations for class$\langle \mathbf{K} \rangle$ and class$\langle \mathbf{K}_0 \rangle$, which would allow further testing of the distribution model hypothesis for the classes;

(2) re-constructing quasi-amino acid composition space in collective amino acid compositions, such as {L,E,A,K,R} and {P,T,V,G} groups (Figure 5) or groups according to hydrophilicity-aromaticity-charge characteristics, that would lower the effective dimension of the geometric space, r, and consequently (a) decrease N($T_v$) and N($T_h$), and the resource requirements for full computational geometry calculations, and (b) increase $\sqrt[r]{N}$, the mean 'linear density' of N points along coordinate axes of an r-dimensional space, and the population density of data distribution in the geometric space;

(3) surveying Protein Data Bank for an optimal set of criteria of secondary structure content, such as 15% of the amino acid residues for α helix and 10% for β strand [5], for robust and automatic classification of the folding type of protein structures that are either experimentally determined [7-9] or computationally predicted [23];

(4) examining the SCOP-CATH mapping [36] in amino acid composition space.

In Summary: Nakashima et al. [5] showed the relationship between amino acid composition and folding type of globular proteins, which at the time appeared to be a promising avenue for protein structure prediction. Sequence-based protein structure prediction methods had been developing at a far greater pace, culminating in the recent and still evolving AlphaFold system [23] and rendering amino acid composition less compelling for practical purposes. From the theoretical biology perspective however the distinctive amino acid composition-protein structure class relationship remains relevant: while methods such as AlphaFold for the most part can answer the question of "what" conformation a polypeptide is folded into, on less solid footing currently are the "how's" and the "why's" of the protein folding process, which in due course could be expected to help where learning-based algorithms may falter and for which any and all fundamental knowledge about protein structures (e.g. the amino acid composition-folding class relationship) would be essential.

A follow up of previous work [6], here we re-cast the problem of amino acid composition-protein structure class problem, and in a 'divide and conquer' strategy focus on data distributions of protein folding classes for deriving quantitative information on their geometry in the high dimensional amino acid composition space. Whereas statistical and machine-learning approaches aim at predictive modeling [16,18,25], the results from modeling full distributions of protein structure classes in the comprehensive hierarchical database SCOP are summary sum total of all knowledge theoretically available, furnishing not only mechanistic understanding of the complex amino acid composition-protein structure class relationship (vis-à-vis any overlaps leading to class ambiguity and classification limits), but also insights into the role of high dimensional amino acid composition space in protein studies.

# Methods

## I. General

A distribution D (with underline) is a set of 20-component vectors $\bar{v}$ (with overline) belonging in a defined group G

$$\underline{D} := \{ \bar{v} \mid \bar{v} \in G \}$$

where $\bar{v}$ is located in the 19D hyper-plane that intersects the 20 coordinate axes at unit distances from the origin, is either a proper amino acid composition vector of fractions of the counts of 20 naturally occurring amino acids to the total residue count of a protein, or one derived from such amino acid composition vectors. A list of properties and geometric quantities of $\underline{D}$ includes

| | |
|---|---|
| $N(\underline{D})$ | cardinality of distribution $\underline{D}$, a set of vectors |
| $P(\underline{D}) := pmmd(\underline{D})$ | ellipsoid derived from product moment matrix diagonalization procedure *pmmd* [6] |
| $M(\underline{D}) := mvee(\underline{D})$ | minimum volume enclosing ellipsoid from procedure *mvee*, python code for arbitrary dimension published in github.com [11] |
| $V_X(\underline{D}) := (\pi^{1/2}/\Gamma(1+r/2))\cdot\Pi r_m(X(\underline{D})),\ \ X \in \{P, M\}$ | volume of ellipsoid X, of rank r and radii $r_m(X(\underline{D}))$ for distribution $\underline{D}$, $\Gamma$ is the gamma function and $\Pi r_m$ is the product of all radii, $r_m$ [6,24] |
| $\dim_M(\underline{D}) := \sqrt[r]{(V_M(\underline{D})/(\pi^{1/2}/\Gamma(1+r/2)))}$ | geometric mean of the radii of mve ellipsoid |

of rank r, as an estimated size per dimension of D suitable for comparison of distributions of different ranks and collective terms,

| | |
|---|---|
| **I** := {1,2,3,4,5} and **J** := {1,2,3,4} | for all classes and four main folding types, respectively, in NNO basis set |
| **K** := {a,b,c,d,g} and **L** := {a,b,c,d} | for classes in SCOP database |
| class⟨**X**⟩ := { class⟨i⟩ \| i ∈ **X** }, **X** ∈ { **I**, **J**, **K**, **L** } | distributions for classes in **X** |
| *M*(**X**) := *M*(class⟨**X**⟩), **X** ∈ { **I**, **J**, **K**, **L** } | mve ellipsoid *M*, and similarly for other properties or quantities |

II. Polytope

(a) Three groups of calculations for a polytope model *T* of distribution D :

1. Redundancy test to identify points that are duplicates, located within the interior of *T*, or otherwise 'redundant' in its definition,

| | |
|---|---|
| redund(D) := {$\bar{v}$ \| $\bar{v}$ ∈ D & $\bar{v}$ ∈ *redundancy*(D)} | set of redundant vectors, itself a distribution, where *redundancy* denotes the 'redund' option of running either program *cdd* in |

cddlib package [13] or *lrs* in lrslib package [15]

and those that are non-redundant,

vertex($\underline{D}$) := $\underline{D}$ - redund($\underline{D}$) — non-redundant or essential vectors

2. Conversions between dual representations of polytope, $T_v$ in vertices and $T_h$ in half-planes, of $\underline{D}$,

$T_h$($\underline{D}$) := half-plane($\underline{D}$) := *h_enum*($\underline{\text{vertex}(D)}$) — dual half-planes, where *h_enum* denotes the main programs *lrs* or *cdd* from their respective packages

$T_v$($\underline{D}$) := vertex($\underline{D}$) := *v_enum*(half-plane($\underline{D}$)) — dual vertices, where *v_enum* is *lrs* or *cdd*

3. For polytope *T* of the distribution $\underline{D}$ of a protein folding class, generally of rank r=19 and embedded in the 19D hyperplane of the amino acid composition space, the content, or volume, is calculated from the vertex set as follows:

$V_T$($\underline{D}$) := $\text{volume}_T$ ($\underline{D}$) — volume of polytope *T* of $\underline{D}$, is the base, $V_{base}$, of the (r+1)D pyramid (apexed at the origin of the coordinate system) according to the formula

$$V_{pyramid}=\text{height}*V_{base}/(r+1)$$

where $V_{pyramid}$ is *volume*($\overline{\text{origin}}$+vertex($\underline{D}$)) computed from running the program *lrs* in lrslib package with the 'volume' option, height is 1/sqrt(r+1), and r is the rank of $\underline{D}$

$V_{T\text{-est}}$($\underline{D}$) := $\text{correl}_r$ ($V_M$($\underline{D}$)) — estimated volume of *T*, where $\text{correl}_r$ is the correlation relationship in Figure 4 for rank r

(b) For geometric relationship between distributions $\underline{D}_1$ and $\underline{D}_2$ :

4. Polytope intersection, or overlap, of $T(\underline{D}_1)$ and $T(\underline{D}_2)$

$\underline{O}_p ( \underline{D}_1, \underline{D}_2 ) := v_enum(T_h(\underline{D}_1)+T_h(\underline{D}_2))$ — vertex enumeration of the combined set of half-planes of polytopes for $\underline{D}_1$ and $\underline{D}_2$ that are full rank (r=19)

5. For centroid overlap $\underline{O}_c( \underline{D}_1 | \underline{D}_2 )$,

$\overline{\text{centroid}}\langle k\rangle := \text{mean}(\underline{\text{Class}}\langle k\rangle)$ — a 20-component vector in the 19D amino acid composition space calculated as the mean of $\overline{v}$'s for vertices of a class distribution, where $\underline{\text{Class}}\langle k\rangle$ is vertex($\underline{\text{class}}\langle k\rangle$) when vertex($\underline{\text{class}}\langle k\rangle$) $\neq$ $\underline{\text{class}}\langle k\rangle$. With redundancies having been excluded, the centroid is a true geometric center, unbiased by any internal structure of the distribution or abundance of highly repeated or closely related sequences

$\underline{O}_c( \underline{D}_1 | \underline{D}_2 ) := \{ \overline{\text{centroid}}(\underline{D}_2) \text{ if redund}( \underline{D}_1 + \overline{\text{centroid}}(\underline{D}_2) ) == \overline{\text{centroid}}(\underline{D}_2) \}$ — a single redundancy test of the vector $\overline{\text{centroid}}$ of $\underline{D}_2$ against distribution $\underline{D}_1$

6. Distribution overlap,

$\underline{O}_{d}(\, \underline{D}_1 \mid \underline{D}_2\, ) := \{\, \bar{v} \mid \bar{v} \in \underline{D}_2$ & $\mathrm{redund}(T_v(\underline{D}_1)+\bar{v}) == \bar{v}\, \}$ N($\underline{D}_2$) single redundancy tests

7. Distribution overlap via stepwise reduction,

$\underline{D}_{2.i+1} := \underline{D}_{2.i} - \{\, \bar{v} \mid \bar{v} \in \underline{D}_{2.i}$ & $\mathrm{vertex}(T_v(\underline{D}_1)+\underline{D}_{2.i})\, \}$, where $\underline{D}_{2.0} = \underline{D}_2$ iterative identification and removal of members of distribution $\underline{D}_2$ that are non-redundant in $\underline{D}_{2.i}$, the combined set of $T_v(\underline{D}_1)$ and $\underline{D}_2$ at a given step i, with the process terminating when $\underline{D}_{2.i}$ is all redundant or is empty. Even for larger sets <u>class</u>⟨**K**⟩, all calculations are completed in at most 6 steps

## Supplementary Material

Supplementary material includes a zip archive, *classes.zip*, of five files in plain text: *class-i*, i=a,b,c,d,g. Each *class-i* file is a list of SCOP families (labeled in ***sccs***'s) constituting the distribution class⟨i⟩. Capitalized ***sccs***'s belong in the corresponding Class⟨i⟩. Each record in *class-i* is identified by an ***sccs*** string, followed by 2 words: (1) amino acid composition in the form of 20 comma-separated fractions, for amino acids in the order of one-letter codes g,a,i,v,l,p,n,k,q,d,e,m,r,c,s,t,h,f,w,y (2) comma-separated SCOP ***sid***'s for the ***sccs***.

## Acknowledgments

The author thanks Komei Fukuda, ETH Zurich, author of cdd package, for information on computational geometry and polyhedron calculations early on, and David Avis, McGill U and Kyoto U, author of lrs package, for help with lrs software and with computational geometry in general. Skip Jordan, Otaru University of Commerce, suggested linear programming solution to class intersections, and kindly shared its procedural details, which was applied to Nakashima dataset. The author also thanks Barry Honig, Biochemistry, Molecular Biophysics and Systems Biology, Columbia U, for critical reading of the manuscript. Open access to other important and essential resources (minimum volume enclosing ellipsoid procedure and SCOP database, etc.) are hereby acknowledged. This is an independent research work, with no academic, governmental, corporate, institutional funding or sponsorship.

## References

[1] Levitt M, Chothia C (1976) Structural patterns in globular proteins. Nature 261:552–558.

[2] Richardson JS (1981) The anatomy and taxonomy of protein structure. Adv Protein Chem 34:167–339.
Updated Web version at https://www.semanticscholar.org/paper/The-anatomy-and-taxonomy-of-protein-structure.-Richardson/d848e0df740036b3c8b10371b4624a1599d39191 and https://dasher.wustl.edu/bio5357/readings/advprotchem-34-167-81.pdf

[3] Davies DR (1964) A correlation between amino acid composition and protein structure. J Mol Biol 9:605-609.

[4] Havsteen BH (1966) A study of the correlation between the amino acid composition and the helical content of proteins. J Mol Biol 10:1-10.

[5] Nakashima H, Nishikawa K, Ooi T (1986) The folding type of a protein is relevant to the amino acid composition. J Biochem 99:153-162.

[6] Mao B, Chou KC, Zhang CT (1994) Protein folding classes: a geometric interpretation of the amino acid composition of globular proteins. Protein Eng 7:319-330.

[7] Orengo CA, Michie AD, Jones S, Jones DT, Swindells MB, Thornton JM (1997) CATH--a hierarchic classification of protein domain structures. Structure 5:1093–1108.

[8] Fox NK, Brenner SE, Chandonia JM (2014) SCOPe: Structural Classification of Proteins—extended, integrating SCOP and ASTRAL data and classification of new structures. Nucleic Acids Res 42:D304-309.

[9] Chandonia JM, Guan L, Lin S, Yu C, Fox NK, Brenner SE (2022) SCOPe: Improvements to the structural classification of proteins—extended database to facilitate variant interpretation and machine learning. Nucleic Acids Research 50:D553–559.

[10] Bowman N, Heath MT (2023) Computing minimum-volume enclosing ellipsoids. Mathematical Programming Computation 15:621-650.

[11] https://gist.github.com/Gabriel-p/4ddd31422a88e7cdf953, based on work by Nima Moshtagh, http://www.mathworks.com/matlabcentral/fileexchange/9542. (download date: 2021-06-21)

[12] Fukuda K, Prodon A, Double description method revisited. In Deza M, Euler R, Manoussakis I, Ed. (1996) Combinatorics and Computer Science, volume 1120 of Lecture Notes in Computer Science, Springer-Verlag, pp 91-111.

[13] Version 0.94m. https://people.inf.ethz.ch/fukudak/cdd_home/ https://github.com/cddlib/cddlib/

[14] Avis D (2000) lrs: A revised implementation of the reverse search vertex enumeration algorithm. In: Polytopes - Combinatorics and Computation, Ed. G. Kalai and G. Ziegler, Birkhauser-Verlag, 177-198.

[15] Version 0.73. https://cgm.cs.mcgill.ca/~avis/C/lrslib/

[16] Eisenhaber F, Frömmel C, Argos P (1996) Prediction of secondary structural content of proteins from their amino acid composition alone. II. The paradox with secondary structural class. Proteins 25:169-179.

[17] Eisenhaber F, Imperiale F, Argos P, Frömmel C (1996) Prediction of secondary structural content of proteins from their amino acid composition alone. I. New analytic vector decomposition methods. Proteins 25:157-168.

[18] Wang ZX, Yuan Z (2000) How good is prediction of protein structural class by the component-coupled method? Proteins 38:165-175.

[19] Nakashima H, Nishikawa K (1994) Discrimination of intracellular and extracellular proteins using amino acid composition and residue-pair frequencies. J Mol Biol 238:54-61.

[20] Chou KC (2001) Prediction of protein cellular attributes using pseudo-amino acid composition. Proteins 43:246-55.

[21] Cedano J, Aloy P, Pérez-Pons JA, Querol E (1997) Relation between amino acid composition and cellular location of proteins. J Mol Biol 266:594-600.

[22] Chou KC (2011) Some remarks on protein attribute prediction and pseudo amino acid composition. J Theor Biol 273:236–247.

[23] Jumper J, Evans R, Pritzel A, Green T, Figurnov M, Ronneberger O, Tunyasuvunakool K, Bates R, Žídek A, Potapenko A, Bridgland A, Meyer C, Kohl SAA, Jain R, Adler J, Back T, Petersen S, Reiman D, Clancy E, Zielinski M, Steinegger M, Pacholska M, Berghammer T, Kohli P, Hassabis D (2021) Highly accurate protein structure prediction with AlphaFold. Nature 596:583-589.

[24] Kendall MG (1961) A course in the geometry of n dimensions, Hafner Pub. Co., New York.

[25] Marrero-Ponce Y, Contreras-Torres E, García-Jacas CR, Barigye SJ, Cubillán N, Alvarado YJ (2015) Novel 3D bio-macromolecular bilinear descriptors for protein science: Predicting protein structural classes. J Theor Biol 374:125-137.

[26] Michie AD, Orengo CA, Thornton JM (1966) Analysis of domain structural class using an automated class assignment protocol. J Mol Biol 262:168-185.

[27] Brenner SE, Chothia C, Hubbard TJP, Murzin AG (1996) Understanding protein structure: Using SCOP for fold interpretation. Chap. 37 in: Doolittle RF, ed. Computer Methods for Macromolecular Sequence Analysis. Methods in Enzymology. Vol. 266. Orlando, FL: Academic Press. 635-643.

[28] Murzin AG, Brenner SE, Hubbard TJP, Chothia C (1995) SCOP: a structural classification of proteins database for the investigation of sequences and structures. J Mol Biol 247:536-540.

[29] https://scop.berkeley.edu/downloads/update/{dir.cla.scope.txt,dir.com.scope.txt} (download date: 2024-04-24)

[30] https://scop.berkeley.edu/downloads/scopeseq-2.08/astral-scopedom-seqres-gd-all-2.08-stable.fa (download date: 2024-04-24)

[31] https://scop.berkeley.edu/astral/seq/ver=2.08&seqOption=0&output=text&id=sid, where sid is the 7-character domain identifier for those that are absent in the astral file (fetch date: 2024-04-30)

[32] Zhang CT, Zhang R (1998) A new quantitative criterion to distinguish between α/β and α+β proteins (domains). FEBS Letters 440:153-157.

[33] Sickmeier M, Hamilton JA, LeGall T, Vacic V, Cortese MS, Tantos A, Szabo B, Tompa P, Chen J, Uversky VN, Obradovic Z, Dunker AK (2007) DisProt: the database of disordered proteins. Nucleic Acids Res 35:D786–D793.

[34] Potenza E, Di Domenico T, Walsh I, Tosatto SCE (2015) MobiDB 2.0: an improved database of intrinsically disordered and mobile proteins. Nucleic Acids Res 43:D315-320.

[35] Naullage PM, Haghighatlari M, Namini A, Teixeira JMC, Li J, Zhang O, Gradinaru CC, Forman-Kay JD, Head-Gordon T (2022) Protein dynamics to define and refine disordered protein ensembles. J Phys Chem B 126:1885-1894.

[36] Csaba G, Birzele F, Zimmer R (2009) Systematic comparison of SCOP and CATH: a new gold standard for protein structure analysis. BMC Struct Biol 9:23.

# Figure Legends

Figure 1

Geometric representations of a data distribution in 2D. The distribution is a point set consisting of 10 vertices (light gray) and 7 interior points (dark gray), and a descriptive model of the distribution is a geometric object that covers the point set. For protein folding classes, each point in the distribution is the endpoint of an amino acid composition vector. In computational geometry calculations [12-15], the convex polytope is defined either as the set of 10 vertices, or as its dual, the set of 10 inequality equations (one of which is explicitly shown as a half-plane in dashed lines). For the three models, pmmd and mve ellipses (ellipsoid and hyper-ellipsoid in higher dimensions) and polytope, the areas (volume and hyper-volume in higher dimensions) are 0.2695, 0.0523, and 0.0353, respectively. Of the three models, removal of the 7 interior points from the distribution changes only the pmmd ellipse, reducing the area to 0.2191, but has no effect on the mve ellipse or the polytope (hence such interior points are 'redundant').

Figure 2

Schematic drawing of pair-wise relationships between polygons (polytopes in 2D). In general when two full-rank polytopes (blue and green) intersect, the overlap is itself a polytope, i.e. the region marked in orange and defined by five vertices (circles) to be enumerated from the combined set of half-planes (dashed lines in blue and green). For polytopes of large number of vertices and half-planes, vertex enumeration is computationally demanding and provisory overlap information can be derived from discrete levels of overlap: (1) full centroid overlap (blue and red polygons), when the polygons are sufficiently close such that each envelope encloses the other centroid; (2) partial centroid overlap (green and red), when only one polytope encloses the other centroid; (3) distribution overlap (blue and green), with or without any centroid overlap, when one polytope encloses some vertices and/or interior points of the other. Formal definitions are found in Methods II(b). Further, polytopes in higher dimensions may yet overlap only at facets (facet overlap, not shown here in 2D schematics), for example two 3D prisms in contact at the long edges. Schematically for class k of SCOP database, *T*(class⟨k⟩) is shown in solid blue lines, two of its Family⟨s⟩ in dotted lines (the families with N of 4 and rank 2, and N of 2 and rank 1 respectively, and their members marked by x), and *T*(class⟨$k_0$⟩) in short dashed lines.

Figure 3

SCOP hierarchy, reproduced with permission, from https://scop.berkeley.edu/help/ver=2.08.

Figure 4

Top: Linear correlation between $V_M$(D) and $V_T$(D) for D=family⟨s⟩ in SCOPe database, N(D)≤51 and rank=19 (222 data points), and D=class⟨**J**⟩ (4 data points). The 95% confidence interval of prediction for the regression is shaded in gray. Bottom: regression coefficients. Enclosed in parentheses are the number of data points for a given rank (columns 1 and 4), and the coefficients and the standard errors in parentheses (columns 2,3, and 5,6).

Figure 5

Comparison of amino acid compositions for folding types α and β. Top panel: Deviations for class⟨1⟩ and class⟨2⟩, computed from the overall centroid of all proteins in NNO basis set. This is comparable to Figure 1 in [5], except for the slightly different Y-scale due to plain, non-normalized composition vectors in this study, and minor differences due to computing in rational number arithmetic. Comparison of distributions between folding classes or between different datasets (namely NNO vs. SCOPe) is simpler and more straightforward in the amino acid composition space of plain amino acid fractions without normalization. Bottom panel: Deviations for $\underline{\text{class}}\langle \text{a}\rangle$ and $\underline{\text{class}}\langle \text{b}\rangle$, computed from $\overline{\text{centroid}}(\overline{\text{centroid}}\langle \mathbf{L}\rangle)$, with a slightly smaller Y-range, and a different horizontal order of amino acids. For completeness the total bar length in unit of standard deviation for each amino acid is noted below the single-letter code.

Figure 6

Depiction of the arrangement of protein structure classes with sizes and inter-class distances generally consistent with those reported in Tables III and IV, only as a visual guide to discussions in the main text. This is not an attempt as a projection, since any geometrical property can be directly calculated from relevant expressions in the full 19-dimensional space. Each class is represented in the figure by a centroid, an ellipse for the class distribution, and an ellipse in thin line for the inner distribution which is about 1/4 the size of full distribution for class b and about 1/3 for classes a, c, and d (Table III). The clustering of the centroids of classes a,c,d and that of b,c,d are shown as back-to-back triangles. The centroid of $\underline{\text{class}}\langle \mathbf{L}\rangle$ is marked by "+" in orange. Ellipses in dot-dashed line schematically represent Nakashima dataset, for which classes a and b do not intersect since there are no feasible solutions for their combined half-plane inequalities (many thanks to Skip Jordan for suggesting linear programming solution and sharing relevant procedural details).

Figure 1. Geometric models (2D data distribution)

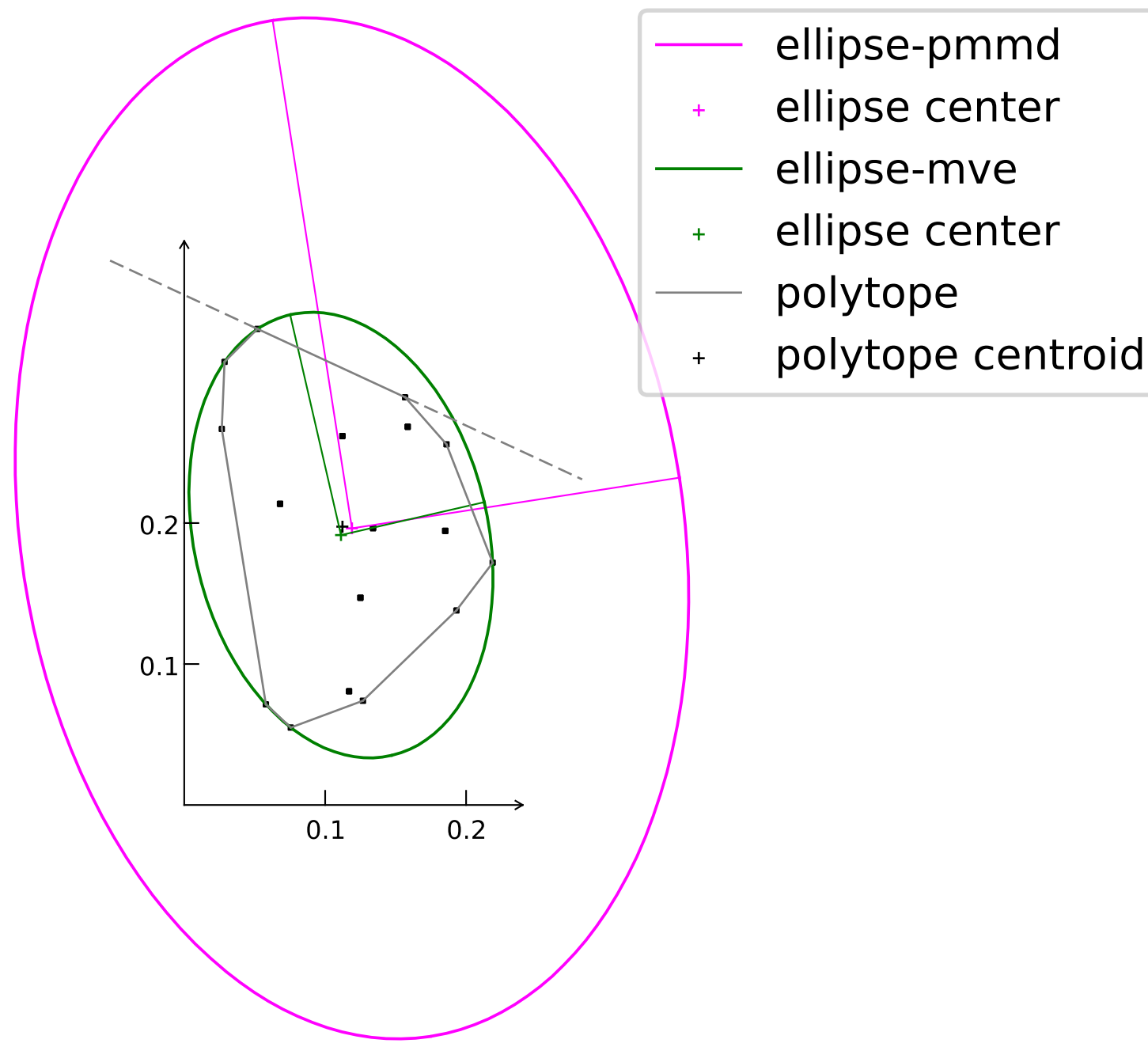

Figure 2. Polytope overlaps

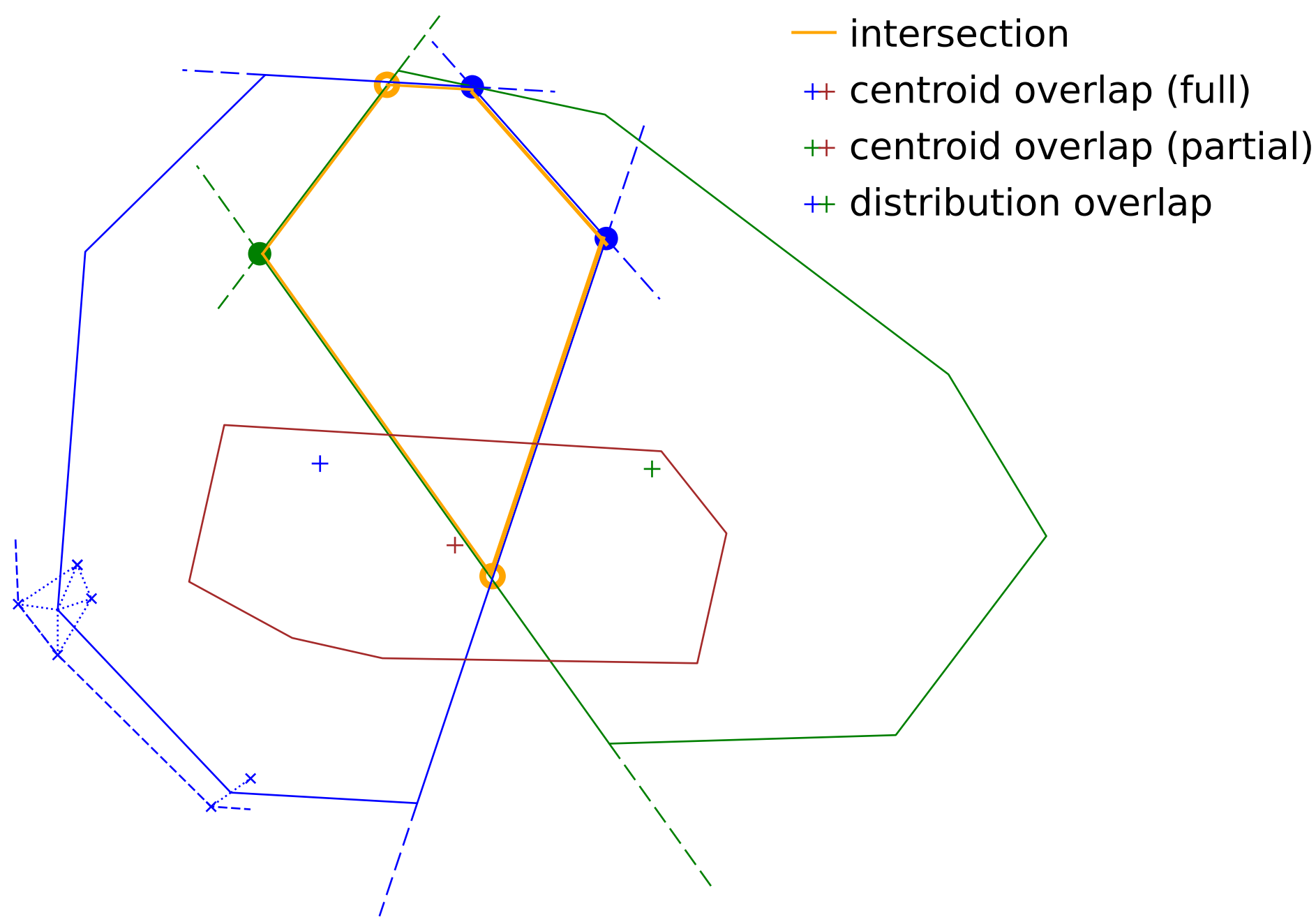

Figure 3. SCOP hierarchy

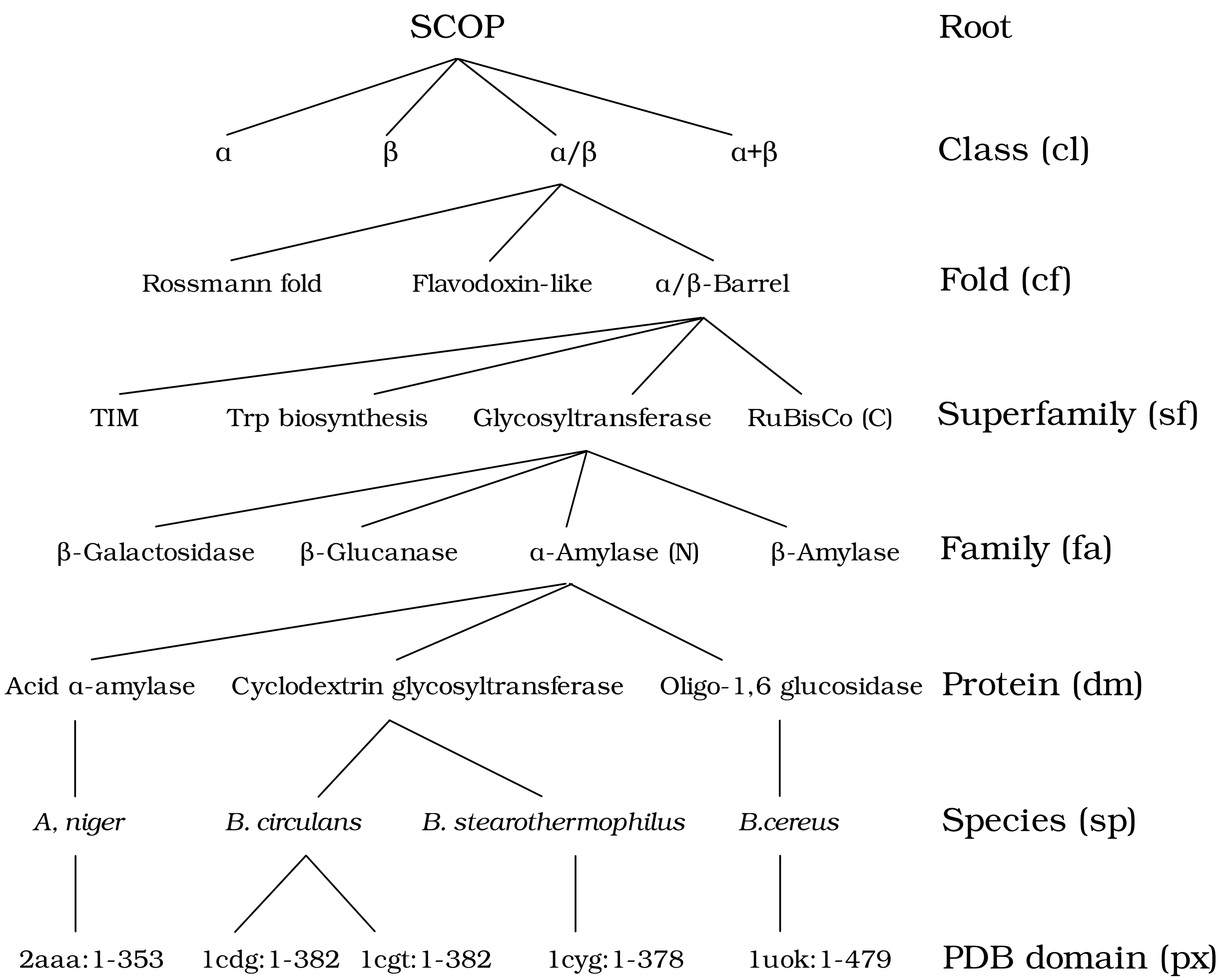

Figure 4. Correlation of $V_M$ and $V_T$

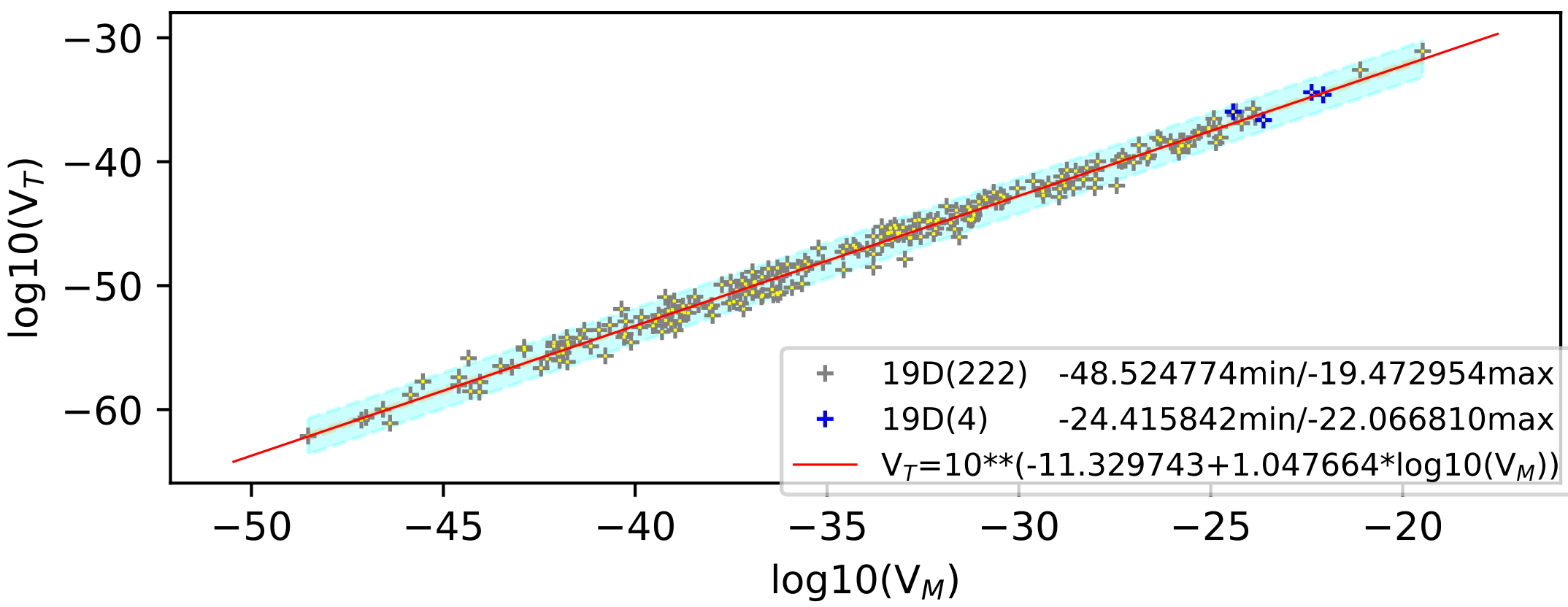

| | | | | | |
|---|---|---|---|---|---|
| 10D( 85) | -10.305(0.859) | 0.868(0.024) | 19D(226) | -11.330(0.778) | 1.048(0.008) |
| 9D( 87) | -10.209(0.910) | 0.822(0.029) | 18D( 36) | -15.280(1.846) | 0.953(0.040) |
| 8D( 97) | -8.893(0.913) | 0.834(0.030) | 17D( 40) | -13.743(1.401) | 0.978(0.031) |
| 7D(122) | -7.444(0.789) | 0.858(0.028) | 16D( 43) | -15.896(1.479) | 0.898(0.034) |
| 6D(134) | -6.586(0.746) | 0.832(0.030) | 15D( 35) | -13.386(1.622) | 0.953(0.041) |
| 5D(183) | -5.712(0.651) | 0.807(0.026) | 14D( 62) | -13.682(1.194) | 0.914(0.029) |
| 4D(251) | -4.651(0.647) | 0.802(0.026) | 13D( 56) | -12.883(1.122) | 0.899(0.028) |
| 3D(261) | -4.000(0.509) | 0.723(0.025) | 12D( 65) | -12.164(1.034) | 0.884(0.028) |
| 2D(362) | -3.175(0.424) | 0.621(0.024) | 11D( 66) | -12.122(0.919) | 0.847(0.026) |

Figure 5. Amino acid compositions, $\alpha$ vs $\beta$

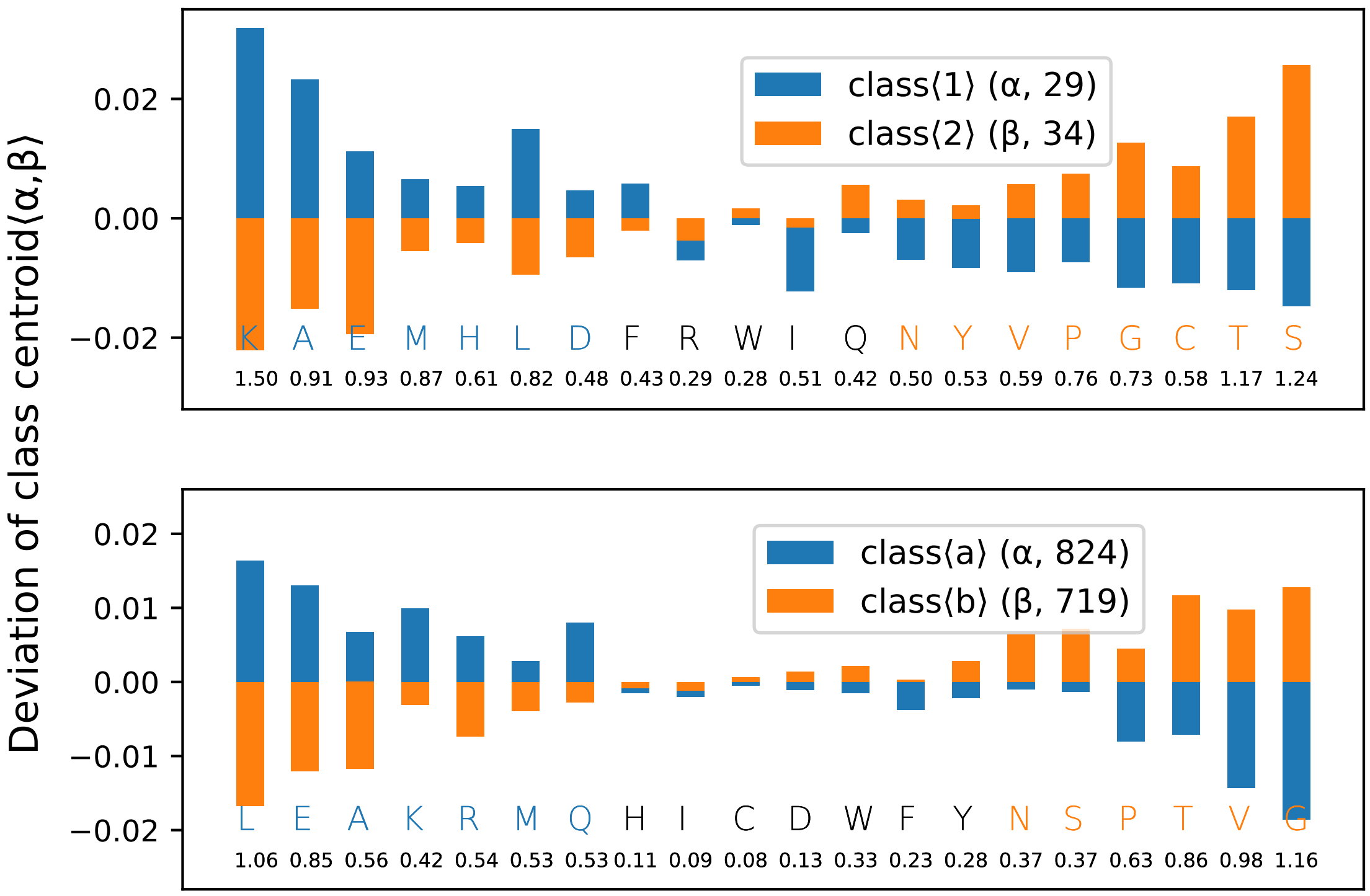

Figure 6. Geometry of protein classes

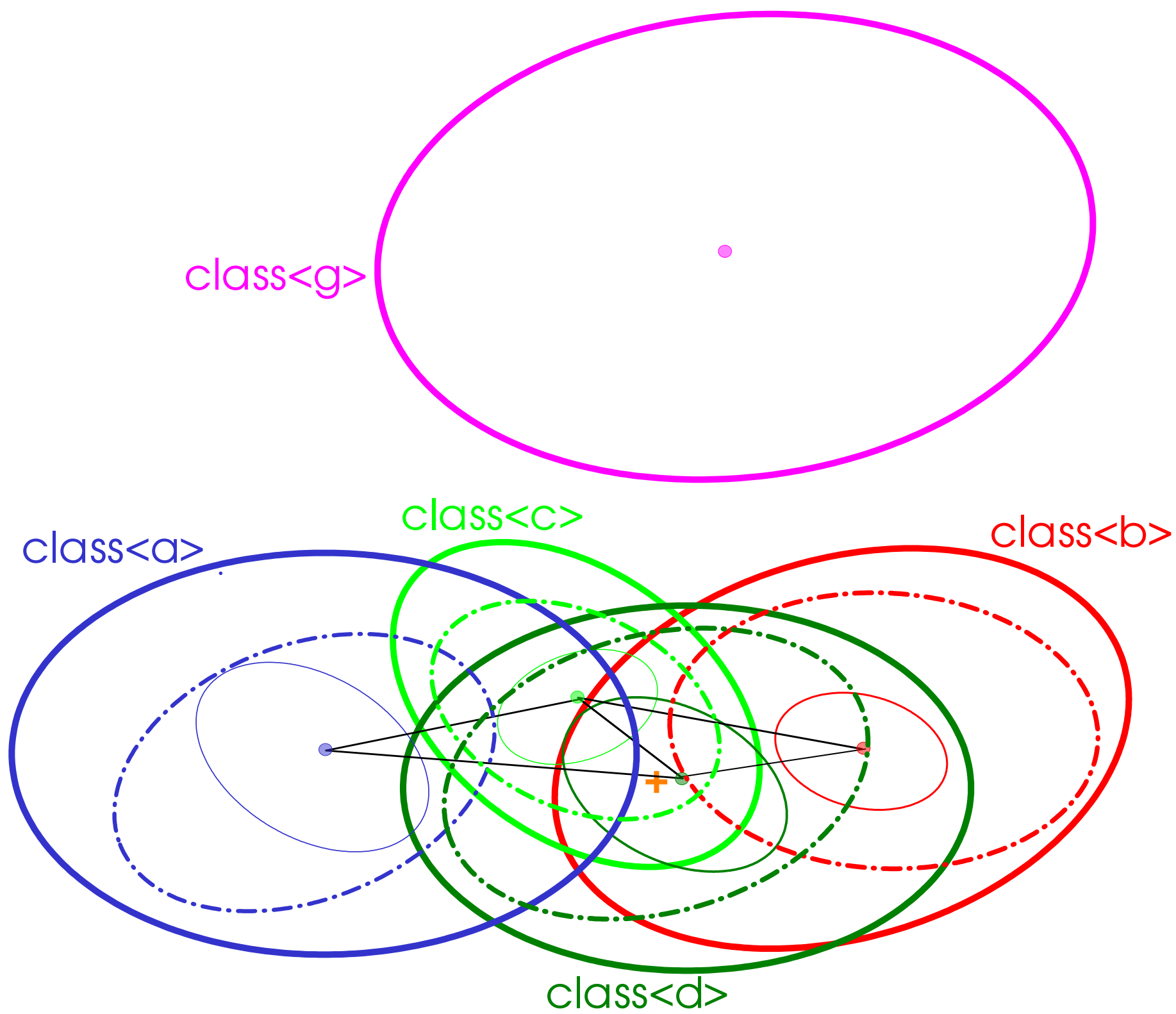

Table I. Geometric parameters of data distributions
NNO basis set of Nakashima et al.

| | protein folding type | α | β | α/β | α+β | (irregular proteins) |
|---|---|---|---|---|---|---|
| | | class⟨1⟩ | class⟨2⟩ | class⟨3⟩ | class⟨4⟩ | class⟨5⟩ |
| | N(class⟨i⟩) | 29 | 34 | 39 | 26 | 4 |
| pmmd ellipsoid* | volume | 3.168e-21 | 4.959e-21 | 1.366e-22 | 3.702e-23 | na |
| mve ellipsoid | volume | 8.574e-23 | 4.257e-23 | 3.838e-25 | 2.369e-24 | na |
| | radii,range | 0.013~0.276 | 0.019~0.246 | 0.019~0.186 | 0.011~0.272 | na |
| | aspect ratio | 20.52 | 18.87 | 9.64 | 24.77 | na |
| polytope | N(vertex⟨i⟩) | 29 | 34 | 39 | 26 | na |
| | N(half-plane⟨i⟩) | 102472 | 1124858 | 6192246 | 18182 | na |
| | volume | 2.455e-35 | 4.003e-35 | 1.059e-36 | 2.256e-37 | na |

The aspect ratio is the simple ratio of longest semi-axis radius to the shortest, as a measure of ellipsoid shape. For class⟨5⟩, there are only 4 proteins, not full-rank and thus without volume and polytope data in the amino acid composition space.

*Although pmmd ellipsoid is deprecated in this study, the volume data are reproduced for backward reference and for comparison here. They are corrected from previously published values which had been calculated from only the product of ellipsoid radii, the third term in expression 14 [6]. The multiplicative factor of 0.0466 (due to first two terms) has been applied for the correction shown here. Values above also differ by a slight increase of precision due to rational arithmetic, and a factor of e-38 due to unit change from percent to simple fraction.

Table II. Pre-processing (a) and statistics (b) of protein structure data from SCOPe database, version 2.08 with 2023-1-6 update

a. **pre-processing**

| | | class(cl) = a~l | cl = a~d,g | cl = a~d,g fa ≠ 0 |
|---|---|---|---|---|
| **dir.cla** | structure domain (px) | 348214 | 290961 | 210541 (196826‡) |
| | protein (dm) | 13526 | 12258 | 11363 (11121‡) |
| | family (fa) | 5433 | 4754 | 3859 (3796‡) |
| px amino acid sequences | | | | **astral**: 208681<br>**wget**: 1860 |
| incomplete or short sequences ignored | | | | Incomplete: 5728<br>23 residues or fewer: 59 |
| | | | | px count final: 204754<br>dm count final: 11100<br>fa count final: 3789 |

b. **statistics**

| | | protein folding type | α | β | α/β | α+β | small proteins |
|---|---|---|---|---|---|---|---|
| | | class | a | b | c | d | g |
| full database | px (total 290961) | | 44921 | 77485 | 84681 | 76691 | 7183 |
| | dm (total 12258) | | 2490 | 2610 | 3085 | 3276 | 797 |
| | fa (total 4754) | | 1089 | 993 | 1003 | 1388* | 281* |
| pre-processing | px (total 204754) | | 32957 | 55705 | 53404 | 56681 | 6007 |
| | dm (total 11100) | | 2285 | 2333 | 2773 | 2966 | 743 |
| | fa (total 3789) | | 893 (895‡) | 766 (777‡) | 807 (826‡) | 1093 (1086‡) | 230 (236‡) |
| | px count,per fa distribution | N(family(s)) | 1~2750 | 1~7754 | 1~1922 | 1~7211 | 1~559 |
| | | N(Family(s)) | 1~528 | 1~2800 | 1~702 | 1~1627 | 1~134 |

**Pre-processing**, proceeding first from left to right, rows 1-3, then top to bottom, column 3. Three streams of data (highlighted in bold) are sourced from SCOPe: **dir.cla** [29], **astral** [30], and **wget** [31]. Class, protein structure domain, protein and family are levels of the SCOPe classification hierarchy, their two-letter code in parentheses in part a. SCOPe family of level '0' (namely SCOP concise classification string of cl.cf.sf.0) are automatic matches and not true families (as noted in the comment file dir.com.scope.txt), and thus excluded. Incomplete sequences are those that contain one-letter amino acid residue code x (for unknown, undetermined, or non-naturally occurring amino acid residues) or X (for sequence discontinuities), for which the amino acid composition vectors are anomalous (the 20 components for the essential amino acid residues do no sum to 1). Lastly, there are a small number of instances of asp/asn (32) and glu/gln (48) ambiguous pairs, in 9 px sequences, labelled with one-letter code of b and z respectively. In each group, half of the instances are assigned to one of the two residues in the pair, and the other half to the second residue.
**Statistics**. Full database, rows1-3, correspond to rows 1-3, column 2 in a.**pre-processing**. Pre-processing, rows 4-6, correspond to rows 9-11, column 3 in a.
‡From file dir.cla.scope.2.07-2020-07-16.txt downloaded on 2020-07-21. Pre-processing did not include short sequences and ambiguous asp/glu pairs.
*Family counts for classes d and g parsed from **dir.cla** are each 1 higher than listed in SCOPe online Statistics.

Table III. Geometric parameters of full and interior data distributions
SCOPe-2.08-2023-01-06

a. **full distribution**

| | protein folding type | α | β | α/β | α+β | (small proteins) |
|---|---|---|---|---|---|---|
| | | class⟨a⟩ | class⟨b⟩ | class⟨c⟩ | class⟨d⟩ | class⟨g⟩ |
| | N(class⟨k⟩) | 893 (895‡) | 766 (777‡) | 807 (826‡) | 1093 (1086‡) | 230 (236‡) |
| mve ellipsoid | volume | 5.523e-18 | 2.118e-18 | 1.975e-21 | 1.418e-18 | 4.721e-17 |
| | radii, range | 0.071~0.324 | 0.060~0.334 | 0.040~0.259 | 0.053~0.284 | 0.073~0.405 |
| | $dim_M$(class⟨k⟩) | 0.145 | 0.137 | 0.096 | 0.135 | 0.162 |
| | aspect ratio | 4.56 | 5.59 | 6.53 | 5.35 | 5.54 |
| polytope | N(Class⟨k⟩) | 824 (819‡) | 719 (723‡) | 744 (753‡) | 993 (981‡) | 229 (228‡) |
| | volume, est. | 3.884e-30 | 1.423e-30 | 9.513e-34 | 9.343e-31 | 3.678e-29 |

b. **interior distribution**

| | N(D) | 69 | 47 | 63 | 100 | 1 |
|---|---|---|---|---|---|---|
| mve ellipsoid | volume | 9.222e-28 | 7.170e-29 | 2.983e-31 | 6.708e-28 | na |
| | radii, range | 0.017~0.117 | 0.013~0.105 | 0.010~0.087 | 0.021~0.106 | na |
| | aspect ratio | 6.86 | 8.01 | 8.89 | 4.99 | na |
| polytope | N(vertex(D)) | 69 | 47 | 63 | 100 | na |
| | volume, est. | 2.216e-40 | 1.525e-41 | 4.885e-44 | 1.588e-40 | na |
| | centroid shift | 0.012 | 0.009 | 0.009 | 0.007 | na |

Estimated polytope volumes (row 7 in a.**full distribution** and row 6 in b.**interior distribution**) were calculated from the regression formula for 19D in Figure 4. The interior distributions for class⟨**L**⟩ are all vertices without redundancy, and for class⟨g⟩, N of 1 and no polytope (rows 1 and 5 in b.). For interior polytopes of classes **L**, the centroid shifts (row 7, **interior distribution**) are calculated relative to centroids of the respective main polytope (those from row 6, **full distribution**). These shifts are small in comparison to the dimensions of the full distributions (rows 3,4, **full distribution**), indicating that (a) the interior polytopes are nearly concentric with the respective full polytopes, and (b) class⟨**L**⟩ distributions are essentially 2-layered geometric structures.
‡Data for SCOPe-2.07-2020-07-16; see Table II for data pre-processing.

# Table IV. Distribution overlap, centroid overlap and distances

a. **Distributiion overlap**

| k' \ k | class⟨a⟩, 893 | class⟨b⟩, 766 | class⟨c⟩, 807 | class⟨d⟩, 1093 | class⟨g⟩, 230 |
|---|---|---|---|---|---|
| Class⟨a⟩, 824 | | 16(10) | 114(70) | 59(12) | 0 |
| Class⟨b⟩, 719 | 4(1) | | 96(52) | 41(4) | 0 |
| Class⟨c⟩, 744 | 2(0) | 3(0) | | 18(1) | 0 |
| Class⟨d⟩, 993 | 25(5) | 52(25) | 195(132) | | 0 |
| Class⟨g⟩, 229 | 0 | 0 | 0 | 0 | |
| Class⟨**K**⟩ | **25(5)** | **55(28)** | **223(160)** | **85(17)** | **0** |

b. **Centroid overlap and distances**

| k' \ k | Class⟨a⟩ | | Class⟨b⟩ | | Class⟨c⟩ | | Class⟨d⟩ | | Class⟨g⟩ | |
|---|---|---|---|---|---|---|---|---|---|---|
| Class⟨a⟩ | | | no | * | yes | * | yes | * | no | * |
| Class⟨b⟩ | no | 0.0696 | | | yes | * | yes | * | no | * |
| Class⟨c⟩ | no | 0.0443 | yes | 0.0391 | | | yes | * | no | * |
| Class⟨d⟩ | yes | 0.0399 | yes | 0.0343 | yes | 0.0171 | | | no | * |
| Class⟨g⟩ | no | 0.1244 | no | 0.1080 | no | 0.1164 | no | 0.1120 | | |
| $\overline{\text{centroid}}$($\overline{\text{centroid}}$⟨**L**⟩) | | 0.0372 | | 0.0343 | | 0.0153 | | 0.0073 | | 0.1122 |
| Class⟨J⟩ | | 0.0687 | | 0.0579 | | 0.0263 | | 0.0552 | | |
| (2.07-2020-07-16) | | 0.0008 | | 0.0011 | | 0.0009 | | 0.0007 | | 0.0013 |

**Distribution overlap**. Rows 1 to 5: $O_d$(Class⟨k'⟩|class⟨k⟩), of class⟨k⟩, labelled at the top, against Class⟨k'⟩, labelled on the left. Labels are followed by the number of families in the class. For each entry of the table the number in parentheses are vertex counts. Row 6: distribution overlap between class⟨k⟩ and Class⟨**K**≠k⟩. None of the overlap is redundant to the matching inner polytope.

**Centroid overlap and distances**. Rows 1 to 5: centroid overlap, yes/no (asymmetric, $O_C$(Class⟨k'⟩|Class⟨k⟩), and centroid-to-centroid distances (symmetric). Row 6: distance of Class⟨k⟩ centroid to the group centroid of four major folding types. NB, distances from group centroid to Class⟨a⟩ and Class⟨b⟩ sum up to be about the distance between Class⟨a⟩ and Class⟨b⟩. Row 7: distance between Class⟨k⟩ and corresponding folding type of NNO basis set. Row 8: distance of Class⟨k⟩ centroid from the corresponding folding type in SCOPe-2.07-2020-07-16 dataset.

*Symmetric distance equal to that from Class⟨k'⟩ to Class⟨k⟩.